# Superconductivity in the layered cage compound $Ba_3Rh_4Ge_{16}$

Yi Zhao[1,2], Jun Deng[3], A. Bhattacharyya[4*], D. T. Adroja[5,6], P. K. Biswas[5], Lingling Gao[1], Weizheng Cao[1], Changhua Li[1], Cuiying Pei[1], Tianping Ying[3,7], Hideo Hosono[7], Yanpeng Qi[1*]

[1] School of Physical Science and Technology, ShanghaiTech University, 393 Middle Huaxia Road, Shanghai 201210, China
[2] University of Chinese Academy of Science, Beijing 100049, China
[3] Beijing National Laboratory for Condensed Matter Physics, Institute of Physics, Chinese Academy of Sciences, Beijing 100190, China
[4] Department of Physics, Ramakrishna Mission Vivekananda Educational and Research Institute, Belur Math, Howrah 711202, West Bengal, India
[5] ISIS Facility, Rutherford Appleton Laboratory, Chilton, Didcot Oxon, OX11 0QX, United Kingdom
[6] Highly Correlated Matter Research Group, Physics Department, University of Johannesburg, PO Box 524, Auckland Park 2006, South Africa
[7] Materials Research Center for Element Strategy, Tokyo Institute of Technology, 4259 Nagatsuta, Midori-ku, Yokohama 226-8503, Japan

* Correspondence should be addressed to Y.Q. (qiyp@shanghaitech.edu.cn) and A.B. (amitava.bhattacharyya@rkmvu.ac.in)

**Abstract** – We report the synthesis and superconducting properties of a layered cage compound $Ba_3Rh_4Ge_{16}$. Similar to $Ba_3Ir_4Ge_{16}$, the compound is composed of 2D networks of cage units, formed by noncubic Rh-Ge building blocks, in marked contrast to the reported rattling compounds. The electrical resistivity, magnetization, specific heat capacity, and μSR measurements unveiled moderately coupled s-wave superconductivity with a critical temperature $T_c$ = 7.0 K, the upper critical field $\mu_0H_{c2}(0)$ ~ 2.5 T, the electron-phonon coupling strength $\lambda_{e-ph}$ ~ 0.80, and the Ginzburg-Landau parameter $\kappa$ ~ 7.89. The mass reduction by the substitution of Ir by Rh is believed to be responsible for the enhancement of $T_c$ and coupling between the cage and guest atoms. Our results highlight the importance of the atomic weight of the framework in cage compounds in controlling the $\lambda_{e-ph}$ strength and $T_c$.

---

**Introduction** – The caged compounds have attracted continuing interest due to their structural flexibility as well as fertile physical properties such as superconductivity. There are two main families of caged superconductors according to the occupation of the guest atom: (i) doped caged compounds represented by fulleride, where the alkali metal is located outside of the soccer-ball shape fullerene (C60)[1-4]. (ii) filled caged compounds, where the guest species sit at the center of the cages. Several representative filled cage superconductors are clathrates ($A_8(Ge,Si)_{46}$, A = alkaline earth)[5-7], β-pyrochlore oxides ($AOs_2O_6$, A = alkali atom)[8-15], filled skutterudites ($MT_4X_{12}$, M = an electropositive cation, T = transition metal, X = pnictogen)[16-25]. and recently discovered metal superhydrides (i.e. $LaH_{10}$, $YH_9$)[26-29].

The emergence of superconductivity in cage compounds is generally considered to be closely related to the interaction between host-guest and interframework[9, 30-33]. So far, much attention has been focused on samples with 3D cubic crystal symmetry. Recently we have synthesized a new cage superconductor $Ba_3Ir_4Ge_{16}$ with $T_c$ = 6.1 K[34]. Interestingly, this compound is composed of 2D networks of cage units, formed by Ir-Ge framework and central Ba atom. The obtained layered structure with a tetragonal lattice (space group, *I4/mmm*) is in marked contrast to its 3D counterparts i.e. filled skutterudites, β-pyrochlore oxides. Low-lying vibration modes contributed by the capsulated Ba cations in $Ba_3Ir_4Ge_{16}$ are responsible for the enhanced electron-phonon coupling strength ($\lambda_{e-ph}$) and appearance of superconductivity.

The discovery of superconductivity in iridium-based caged compounds inspires us to explore novel caged superconductors with other transition metals[34-36]. Herein, we systematically investigate the superconductivity in $Ba_3Rh_4Ge_{16}$, which is the isostructural to the layer cage superconductor $Ba_3Ir_4Ge_{16}$. A systematic study through electrical resistivity, magnetization, heat capacity, and μSR suggested that $Ba_3Rh_4Ge_{16}$ could be categorized as a Bardeen–Cooper–Schrieffer (BCS) superconductor ($T_c$ = 7.0 K) with intermediate coupling. Rh and Ir locate in the same column in the periodic table of elements, however, with different masses. Our results indicate that the weight of the framework in cage compounds provides a feasible tuning knob towards the modulation of $\lambda_{e-ph}$ strength and $T_c$.

**Experiment** – Polycrystalline $Ba_3Rh_4Ge_{16}$ was prepared from stoichiometric amounts of high purity elements by argon arc melting and subsequently annealing in evacuated quartz capsules at 1000°C for 20h. The Powder x-ray diffraction

(PXRD) patterns were collected using a Bruker diffractometer model D8 Advance (Cu rotating anode). The open sources software package FULLPROF was used for Rietveld refinement and the crystal structure was drawn by VESTA. The temperature dependence of dc electrical resistivity was measured at 1.8-300 K by the conventional four-probe method with a physical property measurement system (PPMS, Quantum Design). Magnetization ($M$) measurements were performed using a vibrating sample magnetometer (MPMS, Quantum Design). Specific heat data were obtained by a conventional thermal relaxation method (PPMS, Quantum Design).

The muon spin relaxation/rotation experiments ($\mu$SR) were performed using the MuSR spectrometer at the ISIS Pulsed Neutron and Muon source, UK[37]. The powder sample was mixed with GE-varnish mounted on a silver (99.995 %) sample holder and put in a helium-3 sorption cryostat. The time dependence of the polarization of the implanted muons is given by $P_\mu(t) = G(t)P_\mu(0)$, where the function $G(t)$ corresponds to the $\mu^+$ spin autocorrelation function, which is determined by the distribution of the magnetic field[38]. The time-dependent asymmetry $A(t)$ is a commonly calculated quantity in $\mu$SR experiments which is proportional to $P_\mu(t)$ and is given by $A(t) = \frac{N_F(t) - \alpha N_B(t)}{N_F(t) + \alpha N_B(t)}$, where $N_F(t)$ and $N_B(t)$ are the number of positrons detected in the forward and backward positions respectively, and $\alpha$ is a calibration constant. Time dependence asymmetry spectra in zero-field (ZF) and transverse field (TF) model were collected at different temperatures between 0.3 K and 8.0 K. An active compensation system was used for the zero-field experiments to cancel any stray magnetic fields at the sample position to a level of ~ 0.001 Oe. ZF-$\mu$SR is a nuanced local magnetism probe through the muon spin precession in any internal magnetic fields at the muon sites. In addition to searching for evidence of any magnetic order, ZF-$\mu$SR can also be used to detect the very low spontaneous magnetic fields in the superconducting state associated with time-reversal symmetry (TRS) breaking[39]. In the superconducting mixed state, TF-$\mu$SR experiments were performed in the presence of an applied 300 Oe field well above the lower critical field and well below the upper critical field. The main aims of the present $\mu$SR analysis were to investigate the superconducting pairing mechanism and search for the breaking of TRS in the superconductive state of Ba$_3$Rh$_4$Ge$_{16}$. Using the WiMDA software package[40], all $\mu$SR data were analyzed.

Density functional theory (DFT) calculations employ the projector augmented wave (PAW) method encoded in the Vienna ab initio simulation package (VASP)[41-43]. The projector augmented-wave method is used to describe the wave functions near the core, and the generalized gradient approximation (GGA) within the Perdew-Burke-Ernzerhof (PBE) parametrization is employed as the electron exchange-correlation functional[44]. We relax the lattice constants, and internal atomic positions with the plane-wave cutoff energy of 500 eV and forces are minimized to less than 0.01 eV/Å. The number of K points in the Monkhorst Pack scheme[45] was 5 × 5 × 2 for structure relaxation and 8 × 8 × 2 for self-consistent calculations.

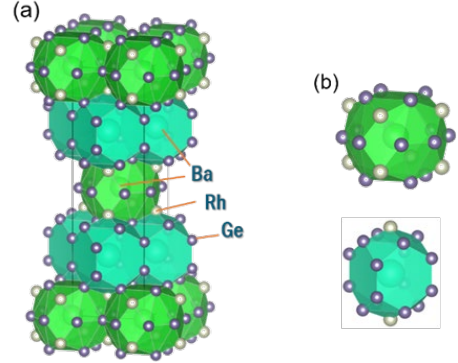

Fig. 1: (a) The crystal structure of Ba$_3$Rh$_4$Ge$_{16}$. (b) the local structure of Ba@[Rh$_8$Ge$_{16}$] (upon) and Ba@[Rh$_2$Ge$_{16}$] (below) cages.

**Results and Discussion –** Fig. 1 shows the crystal structure of Ba$_3$Rh$_4$Ge$_{16}$. The obtained crystal has a tetragonal structure (space group $I4/mmm$, Z = 1). The lattice constants $a$ = 6.5587Å and $c$ = 22.0231Å were consistent with the previous report[36, 46, 47]. Compared with that of Ba$_3$Ir$_4$Ge$_{16}$, the $a$-axis increases while the $c$-axis decreases, leading to an approximately invariant cell volume by the substitution of Ir with Rh. The corresponding crystallographic data are summarized in Table I. The compound is composed of 2D networks with layers made of Rh-Ge cages, which is in sharp contrast to the 3D cage-forming frameworks usually with cubic crystal symmetry. In Ba$_3$Rh$_4$Ge$_{16}$, the [Rh$_8$Ge$_{16}$]$^{2-}$ cages are connected by [Rh$_2$Ge$_{16}$]$^{2-}$ cages encapsulating barium atoms. The crystal structure is a stack of an alternation of Ba@[Rh$_2$Ge$_{16}$] layers and Ba@[Rh$_8$Ge$_{16}$] ones with an interlayer face-sharing connection. Fig. 1(b) shows the isolated Ba@[Rh$_2$Ge$_{16}$] and Ba@[Rh$_8$Ge$_{16}$] building blocks. The shortest Ge -Ge distance (2.50Å) is approximately equivalent to that of cubic Ge, indicating a strong covalent bond. On the other hand, Ba cation is weakly attached to Ge with the shortest Ba-Ge distances of 3.506(2) and 3.452(1) Å in the above two cages. This enhanced space for Ba cations could effectively increase the inharmonic vibration components and thus reduce the rattling frequency of the Ba cations.

Fig. 2(a) shows the temperature dependence of the electrical resistivity for the Ba$_3$Rh$_4$Ge$_{16}$ in the range of 1.8–300 K. The compound exhibits a typical metallic behavior in the whole temperature range and the $\rho(T)$ curve is concave-downwards since electron-electron scattering is dominant in the low-temperature region. A similar trend of resistivity has been observed in its sister compound Ba$_3$Ir$_4$Ge$_{16}$ and also in other caged superconductors, such as β-pyrochlore oxides and filled skutterudites. The $\rho(T)$ data can be well described by the phonon-assisted Bloch-Grüneisen formula:

$$\frac{1}{\rho(T)} = \frac{1}{\rho_i} + \frac{1}{\rho_s} \quad (1)$$

$$\rho_s = \frac{v_F}{\varepsilon_0 \Omega_p^2 a} \quad (2)$$

$$\rho_i(T) = \rho(0) + A\left(\frac{T}{\Theta_R^*}\right)^2 \int_0^{\frac{\Theta_R^*}{T}} \frac{x^2}{(e^x-1)(1-e^{-x})} dx \quad (3)$$

Where $\rho_s$ is the temperature-independent saturation resistivity, $\rho_i$ is the ideal temperature-dependent resistivity dominated by electron-electron scattering, $\rho(0)$ is the residual resistivity, $v_F$ is the Fermi velocity, $\Omega_p$ is the plasma frequency, $a$ is the interatomic spacing, $\Theta_R^*$ is the Bloch-Grüneisen temperature[48-51], and A is any constant depending on the material. The fitting of $\rho(T)$ gives $\Theta_R^* = 237K$. In the normal state (7 K < T < 35 K), the temperature dependence of resistivity could be fitted to a power-law relation:

$$\rho = \rho_0 + aT^n \quad (4)$$

which yields an optimum value for $n$ of 2, indicating a preference for the Fermi liquid behavior.

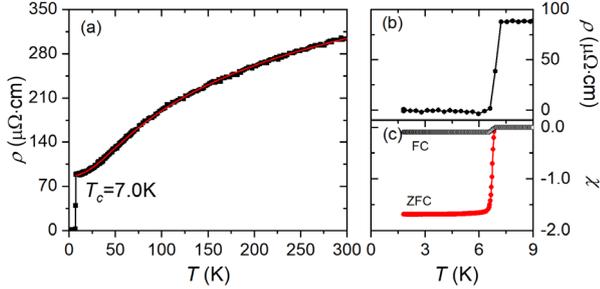

Fig. 2: (a) Temperature-dependent resistivity of $Ba_3Rh_4Ge_{16}$. A red fitting curve is included by using Bloch-Grüneisen formula. (b) Low-temperature enlargement of the measured resistivity under zero magnetic field. (c) Temperature dependence of the magnetic susceptibility measured in zero-field cooling (ZFC) and field cooling (FC) modes under an external magnetic field of 5 Oe.

At low temperatures, a sharp drop in $\rho(T)$ to zero was observed, suggesting the onset of a superconducting transition. Fig. 2(b) shows an enlargement of the superconducting transition. The transition temperatures $T_c$ at 7.0 K is higher than that of $Ba_3Ir_4Ge_{16}$ ($T_c$ = 6.1 K)[34, 36]. The transition widths $\Delta T_c$ approximate 0.3 K, implying fairly good sample quality. In Fig. 2(c), the bulk superconductivity was further confirmed by the large diamagnetic signals and the saturation trend with the decreasing of temperature. As shown in Fig. 2(c), the rapid drop of the zero-field-cooling data denotes the onset of superconductivity, in agreement with temperature of zero resistivity. The difference between zero-field-cooling and field-cooling curves indicates the typical behavior of type-II superconductors. The superconducting volume fractions (SVF) at 1.8 K were larger than 100%, which may be caused by the uncorrected demagnetization factor of the samples.

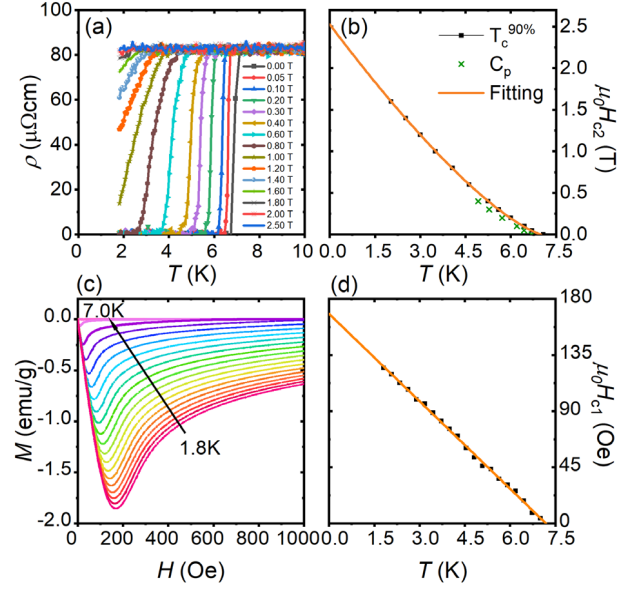

Fig. 3. (a) Temperature dependence of the resistivity within various external fields. (b) The upper critical field $\mu_0H_{c2}$ depends on temperature. The $T_c$ is defined as the temperature at which resistivity drops to 90% of its residual value in the normal state, and green crosses come from heat capacity data. The yellow curve is the best least-squares fit of the equation $H_{c2}(T) = H_{c2}^*(1 - T/T_c)^{1+\alpha}$ to the experimental data. (c) Magnetic hysteresis under various temperatures. (d) The lower critical field $\mu_0H_{c1}$ as a function of temperature. The yellow line is a linear fit.

We conducted resistivity measurements in the vicinity of $T_c$ for various external magnetic fields. As can be seen in Fig. 3(a), the superconducting transition gradually shifted to lower temperatures. At $\mu_0H$ = 2.5 T, the superconducting transition could not be observed above 1.8 K. We plotted the upper critical field ($H_{c2}$) in Fig. 3(b) by using the 90% drop of the resistivity. Deviating from the Wertheimer–Helfand–Hohenberg theory based on the single-band model, the upper critical field of $Ba_3Rh_4Ge_{16}$ shows a positive curvature close to $T_c$. Similar behavior has been observed in $WTe_2$, $MoTe_2$, $MgB_2$, and $NbSe_2$[52-54]. The upper critical field $H_{c2}(0)$ can be described within the entire temperature range by the expression $H_{c2}(T) = H_{c2}^*(1 - T/T_c)^{1+\alpha}$. The fitting parameter $H_{c2}^* = 2.5$ T can be considered as the upper limit for the upper critical field $H_{c2}(0)$. The coherence length ($\xi$) can be acquired from Ginzburg-Landau (GL) equation:

$$\mu_0 H_{c2}(0) = \frac{\Phi_0}{2\pi\xi_0^2} \quad (5)$$

where $\Phi_0$ is the magnetic flux unit, and it can be estimated to be $\xi_0 = 18.1$nm by substituting in $\mu_0H_{c2}(0) = 2.5$T. This estimated value of $H_{c2}(0)$ is well below the Pauli-Clogston limit $\mu_0H_p(0) = 1.84T_c = 12.88$T, lying within the framework of BCS theory.

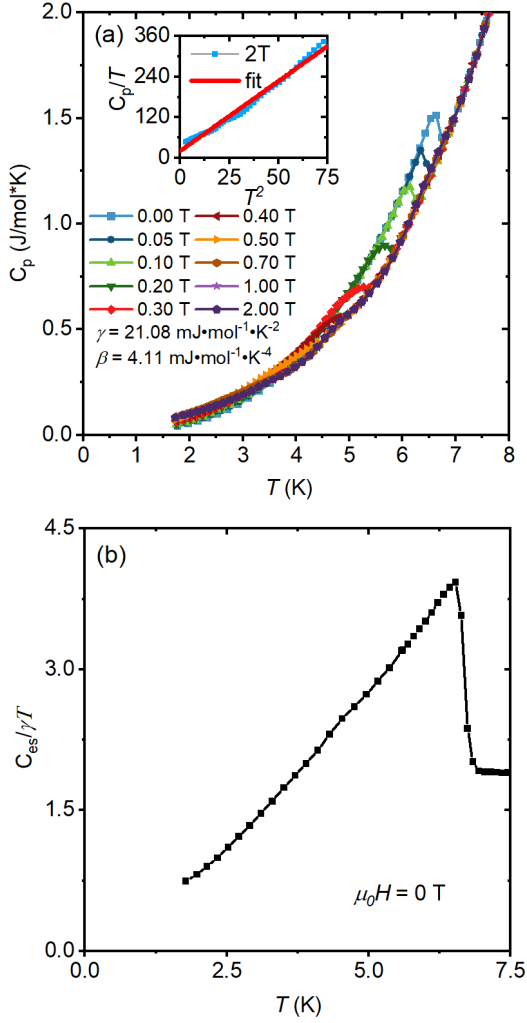

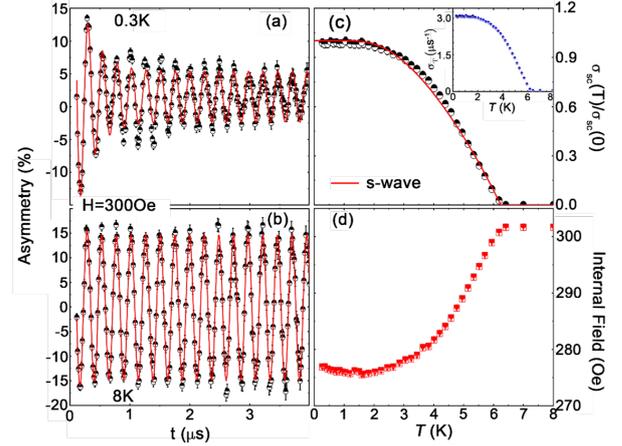

Fig. 5. Transverse field $\mu$SR asymmetry spectra for $Ba_3Rh_4Ge_{16}$ collected (a) at $T = 0.3$ K (b) at $T = 8.0$ K in a magnetic field H = 300 Oe. The line is a fit to the data using equation 12 in the text. (c) The temperature dependence of the normalized superconducting depolarization rate $\sigma_{sc}(T)= \sigma_{sc}(0)$ in presence of an applied magnetic field of 300Oe. The inset shows the temperature variation of the total superconducting depolarization rate $\sigma(T)$. (d) Temperature variation of the internal field.

Fig. 4. (a) The Heat capacity of $Ba_3Rh_4Ge_{16}$ against temperature in different external magnetic fields. (b) Temperature dependence of the electronic specific heat plotted as $C_{es}/\gamma T$ against T in zero field.

Fig. 3(c) shows the low-field magnetization M(H) of $Ba_3Rh_4Ge_{16}$ taken at various temperatures. The linear region at low fields describes the complete Meissner shielding effect. The field that deviates from a linear curve of full Meissner effect was deemed as the lower critical field at each temperature and summarized in Fig.3 (d). A linear fit give $\mu_0 H_{c1}(0) = $ 168 Oe. By using the formula:

$$\mu_0 H_{c1}(0) = \frac{\Phi_0}{4\pi\lambda^2} \ln\left(\frac{\lambda}{\xi}\right) \quad (6)$$

we get the penetration depth $\lambda = 141.96$ nm. The calculated GL parameter of $\kappa = \lambda/\xi \sim 7.89$ confirms the type-II superconductivity in $Ba_3Rh_4Ge_{16}$.

The heat capacities in different magnetic fields were measured by the thermal relaxation method, and the result is shown in Fig. 4(a). The bulk superconductivity is evidenced from a distinct anomaly at 6.8 K. With increasing the magnetic field, the superconducting transition temperature was suppressed. The data were also summarized in Fig. 4(a). As shown inset of Fig. 4(a), a combination of the Debye model, $C_p/T = \gamma + \beta T^2$, was well to fit the curve measured at 2 T. The fitted curves yielded $\gamma = 21.08$ mJ·mol$^{-1}$·K$^{-2}$ and $\beta = 4.11$ mJ·mol$^{-1}$·K$^{-4}$. By using the formula:

$$\beta = N\frac{12\pi^4}{5\Theta_D^3}R \quad (7)$$

and N = 23 for $Ba_3Rh_4Ge_{16}$, we get a Debye temperature of $\Theta_D = 221$ K. The obtained Bloch-Grüneisen temperature is slight larger than the Debye temperature. In addition, $\gamma$ is related to the density of states (DOS) values at Fermi energy $N(E_F)$ by

$$\gamma = \frac{2}{3}\pi^2 k_B^2 N(E_F^0) \quad (8)$$

where $k_B$ is the Boltzmann constant, and $N(E_F^0)$ is density of state for one spin direction and is estimated to be 4.48 states/$eV$ for $Ba_3Rh_4Ge_{16}$. The electronic specific heat ($C_{es}$) was obtained by subtracting the phonon contribution. As shown in Fig. 4(b), the normalized specific heat exhibits a jump around the $T_c$, and the $\Delta C_{es}/\gamma T_c$ is estimated to be 1.77, which is slightly lower than the value of $Ba_3Ir_4Ge_{16}$ (1.85), but is still much larger than the Bardeen-Cooper-Schrieffer (BCS) limit in weak-coupling (1.43). The superconducting gap energy $2\Delta(0)$ can be acquired through,

$$\Delta U(0) = -\frac{\gamma T_c^2}{2} + \int_0^{T_c} C_{es}(T)dT \quad (9)$$

$$\Delta U(0) = \frac{1}{2}N(E_F^0)\Delta^2(0) \quad (10)$$

where $\Delta U(0)$ is the condensation energy $\langle E\rangle_s - \langle E\rangle_n$ and the estimated result of $2\Delta(0)$ is $3.52\ k_B T_c$. According to the McMillan formula for electron-phonon mediated superconductivity[55], $\lambda_{e-ph}$ can be calculated by McMillans relation[55-58]

$$\lambda_{e-ph} = \frac{\mu^* \ln(1.45T_c/\Theta_D) - 1.04}{1.04 + \ln(1.45T_c/\Theta_D)(1-0.62\mu^*)} \quad (11)$$

where $\mu^* = 0.13$ which is a common value for Coulomb pseudopotential[55]. By using the experimental values of $T_c$ and $\Theta_D$, we obtained $\lambda_{e-ph} = 0.80$ for Ba$_3$Rh$_4$Ge$_{16}$ which is similar to the value of Ba$_3$Ir$_4$Ge$_{16}$ (0.75) but the weight of Ba$_3$Rh$_4$Ge$_{16}$ cages are lighter than that of Ba$_3$Ir$_4$Ge$_{16}$, which may be responsible for the enhancement of $T_c$. All of the obtained parameters are listed in Table I.

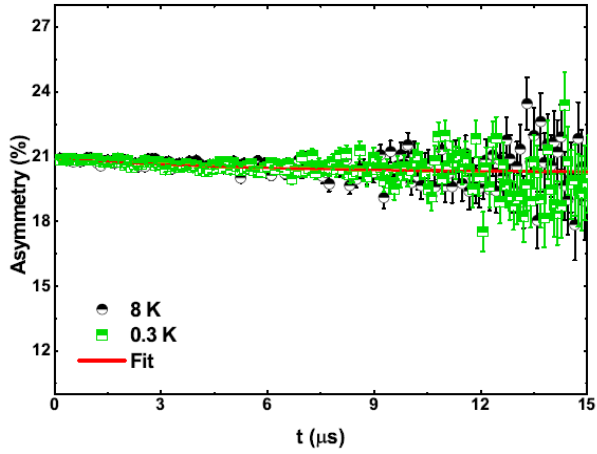

Fig.6. Zero field $\mu$SR time spectra for Ba$_3$Rh$_4$Ge$_{16}$ collected at 0.3 K (green square) and 8 K (black circle) are shown together with lines that are least-square fit to the delta.

To reveal the superconducting gap structure and the pairing symmetry of Ba$_3$Rh$_4$Ge$_{16}$ compound we have carried out TF-$\mu$SR measurements in a field of 300 Oe. The asymmetry spectra of TF-$\mu$SR above and below $T_c$ are displayed in Figs. 5(a)-(b). Because of the inhomogeneous field distribution caused by the existence of the vortex lattice, the asymmetry spectrum decays with time below $T_c$. An oscillating function with a Gaussian decay component is used to fit the time dependence of the TF-$\mu$SR asymmetry spectra at all temperatures:[59-61]

$$G_{z1}(t) = A_1 \cos(\omega_1 t + \phi)\exp\left(-\frac{\sigma^2 t^2}{2}\right) + A_2\cos(\omega_2 t + \phi) \quad (12)$$

where $A_1$= 75.02% and $A_2$= 24.98% are asymmetries of the transverse field and $\omega_1$ and $\omega_2$ come from sample and sample holder frequencies of the muon precession, respectively, and $\phi$ is the phase offset and $\sigma$ is the total Gaussian muon spin relaxation rate. In $\sigma$, there is contribution from the superconducting state ($\sigma_{sc}$) and the normal state $\sigma_n$, which are related by $\sigma^2 = \sigma_{SC}^2 + \sigma_n^2$, here $\sigma_n$ (= 0.0375 $\mu s^{-1}$) is assumed to be constant over the entire temperature range. The fitting of the TF-$\mu$SR data shown in Fig. 5(a)-(b), $A_2$ has been set to 0.2498, which was estimated from fitting the lowest set of data. Although $A_1$ could vary, its value of 0.7502 is almost independent of temperature. The phase value was also calculated using the lowest set of temperature data before being fixed for all other values. The temperature variation of the penetration depth/superfluid density modeled using[61-64].

$$\begin{aligned}\frac{\sigma_{sc}(T)}{\sigma_{sc}(0)} &= \frac{\lambda^{-2}(T,\Delta_{0,i})}{\lambda^{-2}(0,\Delta_{0,i})} \\ &= 1 + \frac{1}{\pi}\int_0^{2\pi}\int_{\Delta(T)}^{\infty}\left(\frac{\delta f}{\delta E}\right) \times \frac{EdEd\phi}{\sqrt{E^2-\Delta(T,\Delta_i)^2}}\end{aligned} \quad (13)^{[62]}$$

where $f = [1 + \exp(E/k_B T)]^{-1}$ is the Fermi function and $\Delta(T,\phi) = \Delta_0\delta(T/T_c)g(\phi)$. The superconducting gap as a function of temperature is approximated by the relation $\delta(T/T_c) = \tanh[1.82[1.018(T_c/T - 1)]^{0.51}]$ where g($\phi$) refers to the angular dependence of the superconducting gap function and $\phi$ is the polar angle for the anisotropy. g($\phi$) is substituted by 1 for an $s$-wave gap[65, 66]. The data can be best described using a single isotropic s-wave gap of 1.00(0.01) meV. This yields a gap of $2\Delta_0/k_B T_C = 3.69$, consistent with the value from heat capacity measurement.

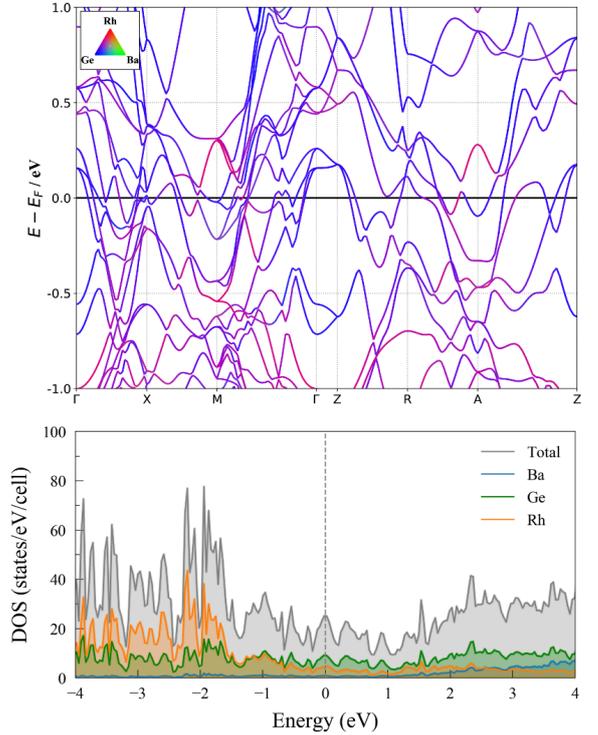

Fig. 7. (a) Band structures along the high-symmetric direction of Ba$_3$Rh$_4$Ge$_{16}$ without SOC. (b) The calculated density of state (DOS) of Ba$_3$Rh$_4$Ge$_{16}$.

To check the presence of any magnetic field ZF-$\mu$SR measurement were performed. The time dependent asymmetry spectra at $T = 0.3$ K ($< T_c$) and 8.0 K ($> T_c$) is depicted in Fig.6. These two spectra shown in the figure are remarkably similar. The lack of muon spin precession eliminates the possibility of an internal magnetic field, as seen in magnetically

ordered compounds. The ZF-$\mu$SR asymmetry spectra is fitted using a Lorentzian function[67]:

$$G_{z2}(t) = A_3\exp(-\lambda t) + A_{\text{bg}}, \quad (14)$$

Here $A_3$ is zero-field asymmetry due to the sample, $A_{\text{bg}}$ is the asymmetry due to the background and $\lambda$ is the relaxation rate which is almost independent of temperature. The asymmetries $A_3$, $A_{\text{bg}}$ are found to be independent of temperature. The solid lines in Fig.6 indicate fits of the ZF-SR asymmetry data using Eq.14, which give $\lambda = 0.2217\ \mu s^{-1}$ at 0.3 K and $\lambda = 0.1850\ \mu s^{-1}$ at 8 K. The values of $\lambda$ at $T \leq T_c$ and $\geq T_c$ coincide within error, indicating that the time reversal symmetry is retained in the superconducting state of $Ba_3Rh_4Ge_{16}$.

Table I: A comparison between $Ba_3Rh_4Ge_{16}$ and $Ba_3Ir_4Ge_{16}$[34]

|  | $Ba_3Rh_4Ge_{16}$ | $Ba_3Ir_4Ge_{16}$ |
| --- | --- | --- |
| a (Å) | 6.5587 | 6.5387 |
| b (Å) | 6.5587 | 6.5387 |
| c (Å) | 22.0231 | 22.2834 |
| $T_c$ (K) | 7.0 | 6.1 |
| $H_{c2}$ (T) | 2.5 |  |
| $H_{c1}$ (Oe) | 168 |  |
| $\gamma$ (mJ mol$^{-1}$K$^{-2}$) | 21.1 | 21 |
| $\Delta C/\gamma T_c$ | 1.77 | 1.85 |
| $\Theta_D$ (K) | 221 | 240 |
| $\xi_0$ (nm) | 18.1 |  |
| $\lambda$ (nm) | 142 |  |
| $\kappa$ | 7.89 |  |
| $\lambda_{e-ph}$ | 0.80 | 0.75 |
| $2\Delta/k_BT_c$ | 3.52 |  |

We also performed DFT calculations to get insight into the superconducting property of $Ba_3Rh_4Ge_{16}$. As shown in Fig. 7a, the hybridization of the electrons from the $p$ orbitals of Ge and $d$ orbitals of Rh dominate around the Fermi level ($E_F$). Ba donates all the 6s$^2$ electrons and its empty bands lie far above $E_F$. Moreover, compared with the DOS of $Ba_3Ir_4Ge_{16}$ [34], the substitution of isoelectric Ir by Rh shifts the local DOS maximum from below $E_F$ to the $E_F$, shown in Fig. 7b. And the N($E_F$) is larger than that of $Ba_3Ir_4Ge_{16}$. This leads to the largest density of states at the Fermi level, and enlarge the electron-phonon coupling in $Ba_3Rh_4Ge_{16}$.

**Summary** – In summary, we systematically investigate superconductivity in a layered caged compound $Ba_3Rh_4Ge_{16}$ with $T_c$ = 7.0 K. Electrical magnetic transport, low-temperature heat capacity, ZF- and TF-$\mu$SR measurements demonstrate that $Ba_3Rh_4Ge_{16}$ could be categorized as a Bardeen–Cooper–Schrieffer superconductor with intermediate coupling. Compare to its isostructural $Ba_3Ir_4Ge_{16}$, replacement of 5$d$-Ir to 4$d$-Rh brings a significant difference towards the weight of the framework in cage compounds, which could control the $\lambda_{e-ph}$ strength. The slight $T_c$ enhancement from 6.1 K to 7.0 K by the isoelectric substitution may originates from the enhancement both $\lambda_{e-ph}$ and DOS at the Fermi level. The present results serve as a useful addition to better understand the rich physics in doped caged materials.

ACKNOWLEDGMENTS

We thank Yuting Wu for valuable discussions. This work was supported the National Key R&D Program of China (Grant No. 2018YFA0704300), the National Natural Science Foundation of China (Grant No. U1932217, 11974246 and 12004252), the Natural Science Foundation of Shanghai (Grant No. 19ZR1477300) and the Science and Technology Commission of Shanghai Municipality (19JC1413900). The authors thank the support from Analytical Instrumentation Center (# SPST-AIC10112914), SPST, ShanghaiTech University. A. Bhattacharyya would like to acknowledge the SERB, India for core research grant support and UK-India Newton funding for funding support. D. T. Adroja would like to thank the Royal Society of London for Newton Advanced Fellowship funding and International Exchange funding between UK and Japan. We thank ISIS Facility for beam time, RB1968041.